\documentclass[aps,pre,reprint,groupedaddress,showpacs]{revtex4-1}

\usepackage{amsmath,amssymb}
\usepackage{graphicx}

\usepackage{color}

\begin{document}

\preprint{}

\title{Real-space renormalization-group approach to the random transverse-field Ising model in finite dimensions}


\author{Ryoji Miyazaki and Hidetoshi Nishimori}
%\thanks{}
\affiliation{Department of Physics, Tokyo Institute of Technology, Oh-okayama, Meguro-ku, Tokyo 152-8551, Japan}


\date{\today}

\begin{abstract}
The transverse-field Ising models with random exchange interactions in finite dimensions are investigated by means of a real-space renormalization-group method.
The scheme yields the exact values of the critical point and critical exponent $\nu$ in one dimension and some previous results in the case of random ferromagnetic interactions are reproduced in two and three dimensions.
We apply the scheme to spin glasses in transverse fields in two and three dimensions, which have not been analyzed very extensively.
The phase diagrams and the critical exponent $\nu$ are obtained, and evidence for the existence of an infinite-randomness fixed point in these models is found.
\end{abstract}

\pacs{05.30.Rt, 75.50.Lk, 64.60.F--, 05.10.Cc}
%\keywords{}

\maketitle

\section{Introduction}{\label{sec:intro}}

Random quantum spin systems have produced fruitful physics beyond our expectations.
One of the turning points that opened the door to this exciting topic was the introduction of random impurities by McCoy and Wu~\cite{McCoy} to the two-dimensional classical Ising model.
The way they introduced the random impurities may seem to be artificial, but using the transfer matrix method, one finds a natural quantum spin system equivalent to the McCoy-Wu model, the random transverse-field Ising spin chain.
After the work by McCoy and Wu, a detailed analysis of the corresponding quantum system revealed interesting and peculiar features of random quantum spin chains~\cite{Fisher}, which are characterized by a so-called infinite-randomness fixed point~\cite{Fisher2}.

Critical phenomena in several low-dimensional random quantum spin systems are considered to be controlled by infinite-randomness fixed points in the renormalization-group picture~\cite{Igloi}.
As a random system with an infinite-randomness fixed point is coarse-grained, the width of distributions grows indefinitely for the logarithms of the parameters in the renormalized Hamiltonian.
In other words, the randomness is infinitely amplified.
The fixed point represents broad distributions of physical quantities.
As a consequence, rare regions in the system, which are usually expected not to affect macroscopic properties, influence some of the behavior of the system.
Average values of physical quantities, in fact, show differences from typical ones.
In addition, a characteristic time scale is also influenced, and the dynamical critical exponent becomes infinity.

Such behavior was discovered in the random transverse-field Ising spin chain through an analytical study~\cite{Fisher}.
The same method has been applied to higher dimensions~\cite{Motrunich, Lin, Karevski, Kovacs0, Kovacs, Kovacs2, Kovacs3} and the same properties have been found consquently in higher-dimensional random ferromagnets.
There is, however, another report with a conflicting result~\cite{Dimitrova} in higher dimensions, and hence further investigations are necessary to resolve the controversy.

The quantum spin-glass model is a particularly important issue among random quantum spin systems, since the properties of critical phenomena in the model have not been clarified very well.
Although the quantum spin-glass model has been presumed to have an infinite-randomness fixed point based on an analogy with the random ferromagnet model~\cite{Motrunich}, the method of the strong-disorder renormalization group~\cite{Igloi} has not successfully settled the issue so far, which is used in most current activities in random quantum spin systems.
In contrast, other numerical estimates~\cite{Guo, Rieger} have indicated results against the conjecture, but the system size used in these studies might be too small.

In the present paper, we study the nature of critical phenomena in the random transverse-field Ising models, more specifically, the random ferromagnetic Ising models and the Ising spin glasses in transverse fields in one, two, and three dimensions, with a method of the real-space renormalization group developed for the non-random models~\cite{Miyazaki}.
Although this method is a variant of the block-spin transformation, a crude approximation in general, our method yields accurate estimates of the critical exponent $\nu$ in non-random systems~\cite{Miyazaki}.
We apply it to models with randomness in this paper.
The one-dimensional model with randomness has been studied with the same strategy as our scheme~\cite{Uzelac}.
The present study covers higher-dimensional cases, spin-glass models in particular.

A special emphasis is placed on resolving the problem of whether or not the infinite-randomness fixed point exists in these models.
Indeed, we have found infinite-randomness fixed points in these models.
The critical exponent $\nu$ for the correlation length has been calculated and its values for the random ferromagnetic Ising model and the Ising spin glass in transverse fields have been found to be very close to each other.
This result suggests that the Ising spin glass in transverse fields might belong to the same universality class as the random ferromagnetic Ising model in transverse fields.

We do not calculate critical exponents other than $\nu$ in the present paper, since our method has not yielded accurate values of the other exponents (for example, $\eta$) in non-random systems~\cite{Miyazaki}.
Also, there are exponents peculiar to the random systems (for example, $\psi$~\cite{Fisher2}) for which we have not established a method to calculate in our framework.
We therefore concentrate on the existence of infinite-randomness fixed points and the exponent $\nu$ in this paper.
It is to be noted that $\nu$ is a representative exponent of infinite-randomness fixed points.

In Sec.~\ref{sec:1D}, we introduce the real-space renormalization-group scheme for the random transverse-field Ising chain after a short review of the model.
Some previous results are correctly reproduced under the scheme.
In particular, the existence of the infinite-randomness fixed point is verified.
Our method is generalized to higher spatial dimensions in Secs.~\ref{sec:2D} and \ref{sec:3D}, where the random ferromagnetic Ising models in transverse fields and Ising spin glasses in transverse fields are investigated.
Our study is concluded in Sec.~\ref{sec:conclusion}.

\section{One dimension}{\label{sec:1D}}

\subsection{Random transverse-field Ising spin chain}
\label{subsec:RTFIC}

Let us recall a few previous results~\cite{Fisher, Pfeuty} related to the random transverse-field Ising chain
\begin{equation}
H = -\sum_{i=1}^{N} J_{i} \sigma_{i}^{z} \sigma_{i+1}^{z} -\sum_{i=1}^{N} \Gamma_{i} \sigma_{i}^{x}, \label{eq:1DTFIM}
\end{equation}
where $\sigma_{i}^{\alpha}$ denotes the $\alpha$-component of the Pauli matrix on site $i$.
The boundary condition is periodic $\sigma_{i} = \sigma_{i+N}$, where the number of spins is $N$, and we assume that $N$ is even.
The couplings $J_{i}$ and transverse fields $\Gamma_{i}$ are random variables independently distributed.
Without loss of generality, we restrict random variables to be positive.
This model has already been analyzed in detail~\cite{Fisher, Pfeuty} and in the following section we reproduce some previously known results with our real-space renormalization-group approach.

When the average of transverse fields is much larger than that of couplings, the system lies in the paramagnetic phase, that is, the expectation value of $\sigma^{z}$ is zero.
A phase transition to the ferromagnetic phase takes place at some point as we reduce the average value of fields.
The system, as a result, obtains a finite expectation value of $\sigma^{z}$.
The para-ferro transition point has been analytically obtained~\cite{Pfeuty},
\begin{equation}
\sum_{i=1}^{N} \ln J_{i} = \sum_{i=1}^{N} \ln \Gamma_{i}, \label{eq:1Dcp}
\end{equation}
which includes the non-random case~\cite{Nishimori}.

The model has an infinite-randomness fixed point.
Critical phenomena controlled by this fixed point are considerably different from the conventional ones.
One of the characteristic features appears in the behavior of the correlation length near the fixed point.  
There are two kinds of critical exponent $\nu$ about the divergence of the correlation length and the two critical exponents take different values in the random transverse-field Ising chain: $\nu = 2$~\cite{Fisher} and $\nu_{\mathrm{typ}} = 1$~\cite{Shankar}.
The first one $\nu$ is for the (average) correlation length $\xi$ and correctly takes into account the effects of randomness.
The other exponent $\nu_{\mathrm{typ}}$ is for the typical correlation length $\xi_{\mathrm{typ}}$, which describes the typical behavior of the system and does not reflect the influence of rare events on macroscopic properties.
The correlation length $\xi$ properly describing the effect of rare events in the random system is defined as the largest length $L$, where the probability that all the spins in the block of length $L$ are correlated, exceeds some finite value~\cite{Chayes, Fisher}.

In the random transverse-field Ising model, the average and variance of $\ln \Gamma - \ln J$ plays an important role in the determination of the correlation-length exponent~\cite{Fisher}.
We now express the average as $\varDelta$,
\begin{equation}
\varDelta = \frac{1}{N} \sum_{i=1}^{N} \ln \frac{\Gamma_{i}}{J_{i}}. \label{eq:deltac}
\end{equation}
According to Eq.~(\ref{eq:1Dcp}), the phase transition in one dimension occurs when this average is equal to zero.
The system lies in the paramagnetic phase if $\varDelta > 0$ and the system in the ferromagnetic phase has a negative $\varDelta$.
The average per block having a length of $L$ is $L \varDelta$.
Roughly speaking, in the Gaussian distribution, for example, the variance of $\ln \Gamma - \ln J$ per block is $L V$, where $V$ expresses the variance per spin,
\begin{equation}
V = \frac{1}{N} \sum_{i=1}^{N} \left( \ln \frac{\Gamma_{i}}{J_{i}} - \varDelta \right)^{2}.
\end{equation}
If $L \varDelta > \sqrt{L V}$, most blocks show paramagnetic behavior, and there are few blocks containing perfectly correlated spins.
Otherwise, there is a significant probability that all spins are correlated in a block.
The relation $\xi \varDelta \sim \sqrt{\xi V}$ is thus a good estimate for the correlation length.
Hence, we have
\begin{equation}
\xi \sim \left( \frac{\varDelta}{\sqrt{V}} \right)^{-2}.
\end{equation}
The correlation length diverges around the transition point $\varDelta = 0$, and we find $\varDelta/ \sqrt{V}$ to be a proper parameter measuring the ``distance" from the critical point.
In general, including the case $\varDelta_{c} \neq 0$, we expect the relation between the correlation length $\xi$ and the critical exponent $\nu$ to be represented as
\begin{equation}
\xi \sim \left( \frac{\varDelta - \varDelta_{c}}{\sqrt{V}} \right)^{- \nu}. \label{eq:xi}
\end{equation}
The above-mentioned crude estimate suggests $\nu = 2$.

If we measure the distance without variance, which reflects the effect of rare events, we can associate the typical correlation length $\xi_{\mathrm{typ}}$ with the critical exponent $\nu_{\mathrm{typ}}$,
\begin{equation}
\xi_{\mathrm{typ}} \sim (\varDelta - \varDelta_{c})^{-\nu_{\mathrm{typ}}}. \label{eq:xityp}
\end{equation}
This expression of the exponent $\nu_{\mathrm{typ}}$ is due to~\cite{Fisher}, a leading study on the one-dimensional model.
Nevertheless, this is not a common definition of $\nu_{\mathrm{typ}}$.
Usually, $\nu_{\mathrm{typ}}$ is expressed as $\nu ( 1 - \psi )$ with another exponent $\psi$ describing the relationship between the length scale and the energy scale~\cite{Fisher2}.
Since we do not have a method to estimate $\psi$, we do not calculate $\nu_{\mathrm{typ}}$ in the present study other than in the simple one-dimensional model.
In higher-dimensional cases, we mention only whether the two exponents defined in Eqs.~(\ref{eq:xi}) and (\ref{eq:xityp}) may have differences.

\subsection{Real-space renormalization group in one dimension}

We develop a real-space renormalization-group procedure for the random transverse-field Ising model in one dimension of Eq.~(\ref{eq:1DTFIM}) at zero temperature~\cite{Uzelac}.
This is a natural generalization of our previous method for the pure transverse-field Ising model~\cite{Miyazaki, Fernandez}.
The method is based on the block-spin transformation preserving the high symmetry of the model.
We can reproduce the exact critical point and critical exponent $\nu$ in the pure transverse-field Ising chain.
This fact is in contrast with standard real-space renormalization-group approaches on the basis of block-spin transformations for quantum systems, which have difficulties in quantitatively accurate estimations (see, e.g.,~\cite{Drell, Jullien, Um, Hirsch, Hu, Fradkin}).

\begin{figure}[b]
  \begin{center}
  \includegraphics[width=4cm,clip]{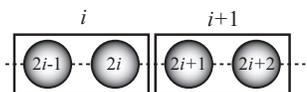}
  \caption{Construction of block spins in one dimension.}
  \label{fig:1Dblocks}
  \end{center}
\end{figure}

We start by dividing the chain into blocks of two spins as shown in Fig.~\ref{fig:1Dblocks}.
The Hamiltonian is also split into intra-block and inter-block parts, 
\begin{gather}
H_{i}^{\mathrm{intra}} = -J_{2i-1} \sigma_{2i-1}^{z} \sigma_{2i}^{z} - \Gamma_{2i-1} \sigma_{2i-1}^{x}, \label{eq:Hintra1D} \\
H_{i,i+1}^{\mathrm{inter}} = -J_{2i} \sigma_{2i}^{z} \sigma_{2i+1}^{z} - \Gamma_{2i} \sigma_{2i}^{x}, \label{eq:Hinter1D}
\end{gather}
where spins $2i-1$ and $2i$ belong to block $i$, and spin $2i+1$ belongs to block $i+1$.
The label of block $i$ runs from $1$ to $N/2$.
Most importantly, this particular block partition is suited to preserve the form of the Hamiltonian under the renormalization-group transformations~\cite{Miyazaki} and is the key for the success of our calculations.

The eigenvalues of $H_{i}^{\mathrm{intra}}$ are degenerate,
\begin{gather}
\varepsilon_{i}^{(1)} = \varepsilon_{i}^{(2)} = - \sqrt{(J_{2i-1})^{2} + (\Gamma_{2i-1})^{2}}, \label{eq:eigenvalue1} \\
\varepsilon_{i}^{(3)} = \varepsilon_{i}^{(4)} = \sqrt{(J_{2i-1})^{2} + (\Gamma_{2i-1})^{2}}.
\end{gather}
The corresponding eigenvectors are
\begin{gather}
|1\rangle_{i} = a_{i}^{+} |\uparrow \uparrow \rangle + a_{i}^{-} |\downarrow \uparrow \rangle, \ \ \ 
|2\rangle_{i} = a_{i}^{+} |\downarrow \downarrow \rangle + a_{i}^{-} |\uparrow \downarrow \rangle, \\
|3\rangle_{i} = a_{i}^{-} |\downarrow \downarrow \rangle - a_{i}^{+} |\uparrow \downarrow \rangle, \ \ \ 
|4 \rangle_{i} = a_{i}^{-} |\uparrow \uparrow \rangle - a_{i}^{+} |\downarrow \uparrow \rangle,
\end{gather}
where
\begin{equation}
a_{i}^{\pm} = \sqrt{\frac{1}{2} \left( 1 \pm \frac{J_{2i-1}}{\sqrt{(J_{2i-1})^{2} + (\Gamma_{2i-1})^{2}}} \right)}, \label{eq:a}
\end{equation}
and $\{|\uparrow \uparrow \rangle, |\uparrow \downarrow \rangle, |\downarrow \uparrow \rangle, |\downarrow \downarrow \rangle \}$ is the orthonormal basis in the $\sigma^z$ basis, i.e., $\sigma^{z}|\uparrow\rangle=|\uparrow\rangle$, $\sigma^{z}|\downarrow\rangle=-|\downarrow\rangle$.

We next keep the two lowest lying energy eigenstates $|1 \rangle$ and $|2 \rangle$, 
and drop the others, $|3 \rangle$ and $|4 \rangle$, to perform a coarse-graining.
This procedure is expected to be effective for the study of the ground state.
We then replace each block with a single spin representing the $|1 \rangle$ and $|2 \rangle$ states.
To this end, we define the projector onto the coarse-grained system as
\begin{equation}
P = \bigotimes_{i = 1}^{N/2} P_{i},
\end{equation}
where $P_{i}$ is the projector,
\begin{equation}
P_{i} = \left( |1\rangle \langle1| + |2\rangle \langle2| \right)_{i}.
\end{equation}
The resulting coarse-grained Hamiltonian is $PHP$.
The renormalized intra-block Hamiltonian is trivially represented by the identity operator $1_{i}$ on block $i$ as
\begin{equation}
P_{i} H_{i}^{\mathrm{intra}} P_{i} = \varepsilon_{i}^{(1)} 1_{i}.
\end{equation}
Terms in the inter-block Hamiltonian are projected as
\begin{equation}
P_{i} \left(1_{2i-1} \otimes \sigma_{2i}^{z} \right) P_{i} = \tilde{\sigma}_{i}^{z}, \label{eq:rsz}
\end{equation}
\begin{equation}
P_{i+1} \left( \sigma_{2i+1}^{z} \otimes 1_{2i+2} \right) P_{i+1} = \frac{J_{2i+1}}{\sqrt{(J_{2i+1})^{2} + (\Gamma_{2i+1})^{2}}} \tilde{\sigma}_{i+1}^{z}, \label{eq:lsz}
\end{equation}
\begin{equation}
P_{i} \left( 1_{2i-1} \otimes \sigma_{2i}^{x} \right) P_{i} = \frac{\Gamma_{2i-1}}{\sqrt{(J_{2i-1})^{2} + (\Gamma_{2i-1})^{2}}} \tilde{\sigma}_{i}^{x}, \label{eq:rsx}
\end{equation}
where $\tilde{\sigma}_{i}^{\alpha}$ is the $\alpha$-component of the Pauli matrix on block $i$, or new site $i$ in the coarse-grained system. 

The renormalized Hamiltonian is consequently expressed as
\begin{equation}
PHP = \sum_{i=1}^{N/2} \varepsilon_{i}^{(1)} 1_{i} - \sum_{i=1}^{N/2} \tilde{J}_{i} \tilde{\sigma}_{i}^{z} \tilde{\sigma}_{i+1}^{z} - \sum_{i=1}^{N/2} \tilde{\Gamma}_{i} \tilde{\sigma}_{i}^{x} \label{eq:1DPHP}
\end{equation}
with renormalized couplings
\begin{equation}
\tilde{J}_{i} = \frac{J_{2i} J_{2i+1}}{\sqrt{(J_{2i+1})^{2} + (\Gamma_{2i+1})^{2}}}, \label{eq:1DrJ}
\end{equation}
\begin{equation}
\tilde{\Gamma}_{i} = \frac{\Gamma_{2i-1} \Gamma_{2i}}{\sqrt{(J_{2i-1})^{2} + (\Gamma_{2i-1})^{2}}}. \label{eq:1DrG}
\end{equation}
Note that our transformation preserves the form of the Hamiltonian.
In other words, our method does not generate additional couplings under renormalization.
Other choices of the intra block and inter block Hamiltonians lead to more inconvenient transformations that do not preserve the form of the Hamiltonian. 

Let us calculate renormalized $\varDelta$ to generate the renormalization-group equation,
\begin{equation}
\begin{split}
\tilde{\varDelta} 
&= \frac{1}{N/2} \sum_{i=1}^{N/2} \ln \frac{\tilde{\Gamma}_{i}}{\tilde{J}_{i}} \\
&= 2 \varDelta,
\end{split}
\end{equation}
where we have used a property of the periodic boundary condition $\tilde{\Gamma}_{N/2 + 1} = \tilde{\Gamma}_{1}$.
The renormalization-group equation $\tilde{\varDelta} = 2\varDelta$ has a fixed point
\begin{equation}
\varDelta_{c} = 0.
\end{equation}
This agrees with the exact transition point~\cite{Pfeuty}.
Combining the change of $\varDelta$ with that of the typical correlation length $\tilde{\xi}_{\mathrm{typ}} = \xi_{\mathrm{typ}}/2$ through the scale transformation with the scaling factor $2$, we obtain the critical exponent 
\begin{equation}
\nu_{\mathrm{typ}} = 1
\end{equation}
under Eq. (\ref{eq:xityp}).
This is also the exact value~\cite{Shankar}.

To take atypical effects into account, we have to explore the change of variance of $\ln \Gamma - \ln J$ through renormalization.
However, it is difficult to analytically investigate it.
We therefore study the change of variance by numerical methods.

In numerical calculations, we first prepare a pool containing $N$ couplings and $N$ transverse fields to construct a chain having $N$ sites, where $N$ has been chosen to be $10^{6}$.
The parameters obey the uniform distributions $p(J) = \theta (J) \theta (1 - J)$ and $p(\Gamma) = \frac{1}{\Gamma_{u}} \theta (\Gamma) \theta (\Gamma_{u} - \Gamma)$, respectively, where if $x >0$, $\theta(x) = 1$, and $\theta(x) =0$ otherwise.
If the upper bound $\Gamma_{u}$ of the values of transverse fields is equal to $1$, the distributions of $J$ and $\Gamma$ coincide, and the system lies at the critical point.

We next perform the renormalization according to Eqs.~(\ref{eq:1DrJ}) and (\ref{eq:1DrG}) with the periodic boundary condition, and the pool is renewed by generating the renormalized couplings and fields.
Then the size of the system becomes a half of that of the original system.
To repeat the renormalization on a large system, we add a copy of all couplings and fields in the renormalized system to the pool.
Consequently, the number of couplings and fields in the pool is recovered and the couplings and fields obey a distribution that is identical to that of the pool before the copies are added.
We then reconstruct a chain with the renormalized parameters in the pool.
In other words, we relabel the parameters to mix originals and copies. 
It is noted that, since $\tilde{J}_{i}$ and $\tilde{\Gamma}_{i+1}$ in the renormalized system share $J_{2i+1}$ and $\Gamma_{2i+1}$ [Eqs.~(\ref{eq:1DrJ}) and (\ref{eq:1DrG})], if $\tilde{J}_{i}$ is relabeled as $\tilde{J}_{j}$, $\tilde{\Gamma}_{i+1}$ has to be relabeled as $\tilde{\Gamma}_{j+1}$. 

Repeating this scheme, we observe the change of the variance of $\ln \Gamma - \ln J$.
To reduce statistical errors, we run up to $100$ samples. 
We set the system very close to the critical point $\varDelta_{c} = 0$ when the calculation is carried out.
More precisely, the upper bound of fields $\Gamma_{u}$ in the initial condition is set equal to $1$, where the distributions of couplings and fields coincide.
Nevertheless these distributions in practice have a small difference owing to the finiteness of the number of couplings and fields.

\begin{figure}
\begin{center}
\includegraphics[width=7cm,clip]{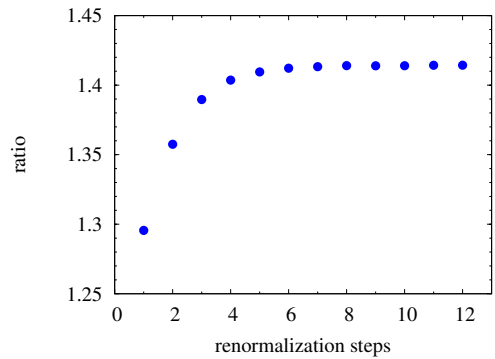}
\caption{Ratio of the square root of the variance of $\ln \Gamma - \ln J$ after a renormalization to that before the renormalization in the random transverse-field Ising model in one dimension.}
\label{fig:1Dvariance}
\end{center}
\end{figure}

A result of the numerical estimate is shown in Fig.~\ref{fig:1Dvariance}, where the ratio of the square root of the variance of $\ln \Gamma - \ln J$ after a renormalization to that before the renormalization is plotted.
The ratio is always larger than $1$, that is, the distributions of the logarithms of parameters keep broadening.
This fact demonstrates the existence of an infinite-randomness fixed point in the random transverse-field Ising spin chain.
Furthermore, the ratio reaches a stationary state after several steps of renormalization.
The value of the critical exponent $\nu$ calculated from Eq.~(\ref{eq:xi}) also becomes stable accordingly because the ratio of $\varDelta$ after a renormalization to that before the renormalization is constant, $2$.
Since randomness is strong when the stationary behavior appears, we expect to extract the nature of the system near the infinite-randomness fixed point from this stationary behavior.
We therefore estimate the critical exponent $\nu$ with the values in the stationary region as 
\begin{equation}
\nu = 2.00(7),
\end{equation}
where we have assumed that a renormalized correlation length is half the original correlation length through the renormalization of scaling factor $2$.
This value is in good agreement with the exact one $\nu = 2$~\cite{Fisher}.

Although our real-space renormalization-group procedure includes approximations, it reproduces the exact critical point, and the exact values of the critical exponents $\nu_{\mathrm{typ}}$ and $\nu$.
Our simple scheme correctly reflects the physics of the infinite-randomness behavior, which is one of the most peculiar features in the random transverse-field Ising spin chain.

\section{Two dimensions}{\label{sec:2D}}

\subsection{Generalization to the two-dimensional models}

\begin{figure}[b]
\begin{center}
\includegraphics[width=7cm,clip]{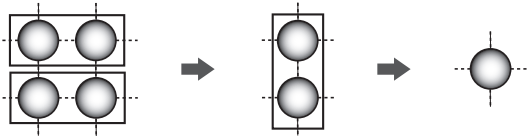}
\caption{Construction of block spins in two dimensions and the two steps of renormalization in the horizontal direction and then the vertical direction.}
\label{fig:2Dblock}
\end{center}
\end{figure}

We generalize the renormalization-group method to the two-dimensional transverse-field Ising model with randomness on the square lattice.
This is also a generalization of the previous study on the pure model in two dimensions~\cite{Miyazaki} to the random model.
The Hamiltonian is
\begin{equation}
H = - \sum_{\langle i, j \rangle} J_{ij} \sigma_{i}^{z} \sigma_{j}^{z} - \Gamma \sum_{i} \sigma_{i}^{x}, \label{eq:TFIM}
\end{equation}
where spins interact with their nearest neighbors $\langle i, j \rangle$.
The key idea consists in performing renormalization-group transformations that preserve the form of the Hamiltonian by a projective isometry that preserves the bond algebra (i.e., the algebra realized by the operators $\sigma_{i}^{z} \sigma_{j}^{z}$ and $\sigma_{i}^{x}$).
Using our experience in one dimension, we divide the lattice into blocks just as in one dimension (Fig.~\ref{fig:2Dblock}).
Furthermore, we combine the one-dimensional block method in horizontal and vertical directions 
to restore the symmetry of the lattice.
Specifically, we iterate the renormalization in two directions: first in the horizontal direction and then in the vertical direction (Fig.~\ref{fig:2Dblock}).

Now we redefine the coupling constants for the horizontal direction and the vertical direction to distinguish these two quantities in this scheme.
The coupling constants between the spin at $(i,j)$ and the neighboring spin to the right side is $J_{h(i,j)}$ and that between the spin at $(i,j)$ and the neighboring spin to the upper side is $J_{v(i,j)}$, where $(i,j)$ denotes the location of a single site on the two-dimensional lattice.

In the first step of the renormalization (in the horizontal direction) we replace each block with a single spin using the same procedure as in the one-dimensional case.
We have the relations corresponding to Eqs.~(\ref{eq:rsz})--(\ref{eq:rsx}),
\begin{equation}
\tilde{P}_{(i,j)} \left(1_{(2i-1,j)} \otimes \sigma_{(2i,j)}^{z} \right) \tilde{P}_{(i,j)} = \tilde{\sigma}_{(i,j)}^{z},
\end{equation}
\begin{equation}
\begin{split}
\tilde{P}_{(i+1,j)}& \left( \sigma_{(2i+1,j)}^{z} \otimes 1_{(2i+2,j)} \right) \tilde{P}_{(i+1,j)} \\
&= \frac{J_{h(2i+1,j)}}{\sqrt{J_{h(2i+1,j)}^{2} + \Gamma_{(2i+1,j)}^{2}}} \tilde{\sigma}_{(i+1,j)}^{z},
\end{split}
\end{equation}
\begin{equation}
\begin{split}
\tilde{P}_{(i,j)}& \left( 1_{(2i-1,j)} \otimes \sigma_{(2i,j)}^{x} \right) \tilde{P}_{(i,j)} \\
&= \frac{\Gamma_{(2i-1,j)}}{\sqrt{J_{h(2i-1,j)}^{2} + \Gamma_{(2i-1,j)}^{2}}} \tilde{\sigma}_{(i,j)}^{x},
\end{split}
\end{equation}
where $\tilde{P}_{(i,j)}$ denotes the projector onto the state space of the block spin, namely, the spin at $(i,j)$ in the renormalized system.
We find that the $z$ component of the spin on the right spot in a block becomes the $z$ component of the block spin, but the $z$ component of the spin on the left spot in a block becomes the $z$ component of the block spin multiplied by $J_{h}/ \sqrt{J_{h}^{2} + \Gamma^{2}}$.
The renormalized couplings and fields are then written as,
\begin{equation}
\tilde{J}_{h(i,j)} = \frac{J_{h(2i,j)} J_{h(2i+1,j)}}{\sqrt{J_{h(2i+1,j)}^{2} + \Gamma_{(2i+1,j)}^{2}}}, \label{eq:r1Jh}
\end{equation}
\begin{equation}
\begin{split}
\tilde{J}_{v(i,j)} =& \frac{J_{h(2i-1,j)}}{\sqrt{J_{h(2i-1,j)}^{2} + \Gamma_{(2i-1,j)}^{2}}} \\
& \times \frac{J_{h(2i-1,j+1)}}{\sqrt{J_{h(2i-1,j+1)}^{2} + \Gamma_{(2i-1,j+1)}^{2}}} \\
& \times J_{v(2i-1,j)} + J_{v(2i,j)}, \label{eq:r1Jv}
\end{split}
\end{equation}
\begin{equation}
\tilde{\Gamma}_{(i,j)} = \frac{\Gamma_{(2i-1,j)} \Gamma_{(2i,j)}}{\sqrt{J_{h(2i-1,j)}^{2} + \Gamma_{(2i-1,j)}^{2}}}.
\end{equation}
In Eq.~(\ref{eq:r1Jv}), $J_{v(2i,j)}$ is derived from the coupling of two spins on the right spot in each block, and the rest is derived from the one on the left spot in the blocks.

Next the system is renormalized in the vertical direction in the same way as the horizontal direction.
The coupling constants and the transverse field are now
\begin{equation}
\begin{split}
\tilde{\tilde{J}}_{h(i,j)} =& \frac{\tilde{J}_{v(i,2j-1)}}{\sqrt{\tilde{J}_{v(i,2j-1)}^{2} + \tilde{\Gamma}_{(i,2j-1)}^{2}}} \\
& \times \frac{\tilde{J}_{v(i+1,2j-1)}}{\sqrt{\tilde{J}_{v(i+1,2j-1)}^{2} + \tilde{\Gamma}_{(i+1,2j-1)}^{2}}} \\
& \times \tilde{J}_{h(i,2j-1)} + \tilde{J}_{h(i,2j)},
\end{split}
\end{equation}
\begin{equation}
\tilde{\tilde{J}}_{v(i,j)} = \frac{\tilde{J}_{v(i,2j)} \tilde{J}_{v(i,2j+1)}}{\sqrt{\tilde{J}_{v(i,2j+1)}^{2} +\tilde{\Gamma}_{(i,2j+1)}^{2}}},
\end{equation}
\begin{equation}
\tilde{\tilde{\Gamma}}_{(i,j)} = \frac{\tilde{\Gamma}_{(i,2j-1)} \tilde{\Gamma}_{(i,2j)}}{\sqrt{\tilde{J}_{v(i,2j-1)}^{2} + \tilde{\Gamma}_{(i,2j-1)}^{2}}}. \label{eq:r2G}
\end{equation}
It is important that our transformations in two dimensions also do not generate extra terms as in the one-dimensional case, and the form of the Hamiltonian is preserved.
In addition, the lattice structure is preserved.

Note that our transformations are not local.
After the first step of the renormalization, $\tilde{J}_{h(i,j)}$, $\tilde{J}_{v(i+1,j)}$, and $\tilde{\Gamma}_{(i+1,j)}$ share $J_{h(2i+1,j)}$ and $\Gamma_{(2i+1,j)}$.
Hence, $\tilde{J}_{h(i,j)}$, $\tilde{J}_{v(i+1,j)}$, and $\tilde{\Gamma}_{(i+1,j)}$ are correlated with each other.
Similarly, $\tilde{J}_{h(i,j+1)}$, $\tilde{J}_{v(i+1,j)}$, and $\tilde{\Gamma}_{(i+1,j+1)}$ share $J_{h(2i+1,j+1)}$ and $\Gamma_{(2i+1,j+1)}$.
Thus, $\tilde{J}_{h(i,j)}$, $\tilde{J}_{h(i,j+1)}$, $\tilde{J}_{v(i+1,j)}$, $\tilde{\Gamma}_{(i+1,j)}$, and $\tilde{\Gamma}_{(i+1,j+1)}$ are correlated with each other.
In a column, couplings in the vertical direction, couplings to the left-hand spin, and fields are mutually correlated after the horizontal renormalization.
The vertical renormalization, accordingly, makes couplings in the horizontal direction, couplings to the lower-side spin, and fields in a row mutually correlate.
Consequently, after two steps of the procedure, no sets of couplings and fields are independent from the others.

\begin{figure}[b]
\begin{center}
\includegraphics[width=8cm,clip]{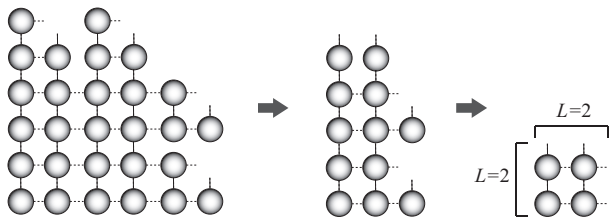}
\caption{Construction of a cluster (left) for the renormalized cluster of size $L = 2$ (right) in two dimensions and two steps of renormalization of the cluster in the horizontal and vertical directions.}
\label{fig:2Dcluster}
\end{center}
\end{figure}

The renormalization-group transformations are numerically performed as follows.
We prepare a pool of couplings $J$ and fields $\Gamma$.
The pool contains $N$ couplings in the horizontal and vertical directions, respectively, as well as $N$ fields.
In our calculation $N$ is $10^{6}$. 
We build a two-dimensional partial lattice, which we call a cluster hereafter, of couplings and fields randomly taken from the pool.
The cluster is renormalized according to Eqs.~(\ref{eq:r1Jh})--(\ref{eq:r2G}) and we put the renormalized cluster in a new pool.
We prepare the shape of the cluster before renormalization to obtain a renormalized square cluster of size $L \times L$.
A renormalization process for a cluster in the case of $L = 2$ for example is depicted in Fig.~\ref{fig:2Dcluster}.
The scheme, from building a cluster to putting the renormalized cluster in the new pool, for a square cluster of size $L \times L$ is executed $N/L \times L$ times.
The new pool is filled by $N/L \times L$ renormalized clusters of size $L \times L$ as a result, and a single renormalization-group transformation for the whole system is completed.
In the next renormalization step, we construct new clusters of the clusters in the pool and repeat the process.
In our calculation $L$ is set equal to $20$.
The renormalization for the whole system is repeated $15$ times, and we run up to $100$ samples.
The procedure is partially based on the calculation method by Nobre~\cite{Nobre} in which we randomly take couplings and fields from the pool as often as we make a cluster.

The reason why we use the clusters in our calculation lies in the nonlocal property of our renormalization-group transformations.
Although in one dimension we have rearranged the labels of couplings and fields to repeat the renormalization calculations on the large system, we cannot freely do so in two dimensions because the renormalized parameters are mutually correlated.
It is difficult to fully keep the correlation in the calculation, but the cluster procedure takes it into account to a certain extent.
We correctly deal with the correlation in clusters.
Although this procedure still ignores the correlation between the clusters on interfaces, the effect of the surfaces is expected to be small if $L$ is large.

\subsection{Random ferromagnet in two dimensions}

We apply the renormalization-group method to the random ferromagnetic Ising model in transverse fields in two dimensions.
The couplings $J$ in the pool are uniformly distributed: $p(J) = \theta (J) \theta (1 - J)$, and the initial value of  the field $\Gamma$ is fixed to a constant at all sites.

We determine the critical point in terms of the parameter $\ln \Gamma - \ln J$ as in the one-dimensional case.
If the average of $\ln \Gamma - \ln J$, which is denoted by $\varDelta$, after $15$ renormalization-group transformations is larger than the initial value of $\varDelta$, we conclude the system to be in the paramagnetic phase.
Otherwise, the system is regarded to be in the ferromagnetic phase.
We obtain the critical value of the transverse field $\Gamma_{c} = 0.9115(5)$ as a result.
This value is to be compared with another estimate $0.84338(2)$~\cite{Kovacs}, which is fairly close to our result in consideration of the simplicity of our idea.

\begin{figure}
\begin{center}
\includegraphics[width=7.5cm,clip]{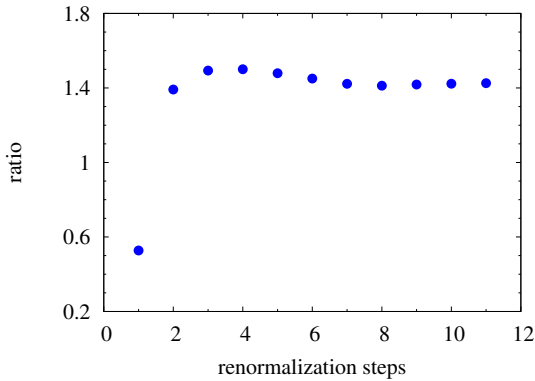}
\caption{Ratio of the square root of the variance of $\ln \Gamma - \ln J$ after a renormalization to that before the renormalization in the random ferromagnetic Ising model in transverse fields in two dimensions.
The initial transverse field is set to be $0.920$.}
\label{fig:2Dvariance}
\end{center}
\end{figure}

We next observe the change of the variance $V$ of $\ln \Gamma - \ln J$.
The ratio of $\sqrt{V}$ after a renormalization to that before the renormalization is plotted in Fig.~\ref{fig:2Dvariance}.
Although the result shows a difference form the result in one dimension in that the ratio in the first renormalization is smaller than $1$, the ratio is always larger than $1$ after that.
We conclude that the two-dimensional random ferromagnetic Ising model in transverse fields has an infinite-randomness fixed point.
This result is the same as in the strong disorder renormalization group~\cite{Motrunich} and Monte Carlo simulation~\cite{Pich}.

To estimate the critical exponent $\nu$ for the average correlation length, we use the relation
\begin{equation}
\frac{\varDelta^{i+1} - \varDelta_{c}^{i+1}}{\sqrt{V^{i+1}}} = 2^{1/\nu_{\mathrm{r}}} \frac{\varDelta^{i} - \varDelta_{c}^{i}}{\sqrt{V^{i}}}. \label{eq:2Dnurelation}
\end{equation}
We call $\nu_{\mathrm{r}}$ the running exponent, which is expected to correspond to the critical exponent $\nu$ if the system is sufficiently close to the infinite-randomness fixed point.
We have assumed that the renormalized correlation length ought to be half the original correlation length through the renormalization with scaling factor $2$.
The symbols $\varDelta^{i}$ and $V^{i}$ denote $\varDelta$ and $V$ renormalized $i$ times, respectively, and $\varDelta_{c}^{i}$ means $\varDelta$ at the transition point after $i$ times of the renormalization-group transformation.
We regard the average of the values of $\varDelta^{i}$ for the initial transverse fields $\Gamma = 0.911$ and $0.912$ as $\varDelta_{c}^{i}$ owing to the uncertainty of our estimation of the transition point.

The value of $\varDelta$ at the transition point depends on the distribution of couplings $J$.
In general the distribution changes through the renormalization-group transformations even if the system lies just on the transition point.
Hence, the value of $\varDelta$ at the transition point after $i$ transformations can be different from the one in the initial distribution.
Moreover, note that the running exponent $\nu_{\mathrm{r}}$ calculated with Eq.~(\ref{eq:2Dnurelation}) can vary with the number $i$ of transformations.

\begin{figure}
\begin{center}
\includegraphics[width=7cm,clip]{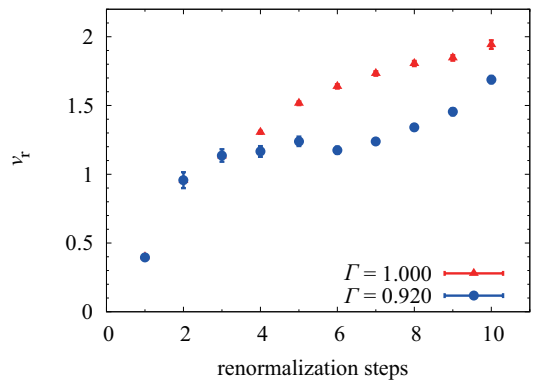}
\caption{Running exponent $\nu_{\mathrm{r}}$ calculated with Eq.~(\ref{eq:2Dnurelation}) in the random ferromagnetic Ising model in transverse fields in two dimensions.
The result in the case $\Gamma = 0.920$, where the system is close to the critical point, has a plateau.}
\label{fig:2Dnu}
\end{center}
\end{figure}

Since we estimate the critical exponent $\nu$ from the running exponent $\nu_{\mathrm{r}}$ near an infinite-randomness fixed point, $\nu_{\mathrm{r}}$ should be calculated near the transition point.
However, instabilities occur if we try to evaluate it by starting too closely to the transition point due to statistical uncertainties.
We thus evaluate $\nu_{\mathrm{r}}$ with the initial transverse field $\Gamma = 0.920$.
The results are shown in Fig.~\ref{fig:2Dnu}.
For comparison, the result of the case of the initial transverse field $\Gamma = 1.000$ is also plotted there. 

We can find a plateau around $4$--$7$ for the result of $\Gamma = 0.920$, whereas it does not exist in the case of $\Gamma = 1.000$.
This plateau is comparable with the stable behavior in the one dimensional case.
We interpret it as a sign of the appearance of a critical phenomenon in our renormalization-group scheme.
The stable behavior in two dimensions does not continue long owing to the difficulty of keeping the system close to the critical point due to randomness, whereas we can easily do so in one dimension because the fixed point in one dimension is fortunately determined only by the average of $\ln \Gamma - \ln J$ and is independent of the distribution of $J$.
This means that in one dimension we have to control only the initial value of the average to prevent the system from migrating away from the fixed point.
Estimating $\nu$ from values of $\nu_{\mathrm{r}}$ on the plateau (from 4 to 7 on the horizontal axis), we obtain 
\begin{equation}
\nu = 1.20(6), \label{eq:2Dnu}
\end{equation}
which is consistent with a previous study from a different approach, namely, the strong-disorder renormalization group, $\nu = 1.24(2)$~\cite{Kovacs}.

Although we do not explicitly estimate the value of $\nu_{\mathrm{typ}}$ defined in Eq.~(\ref{eq:xityp}), which is slightly different from the conventional definition for the reason mentioned in Sec.~\ref{subsec:RTFIC}, we recognize a difference between the values of $\nu$ and $\nu_{\mathrm{typ}}$.
Since the variance $V$ of $\ln \Gamma - \ln J$ keeps growing through the renormalization-group transformations, the change of $(\varDelta - \varDelta_{c})/\sqrt{V}$ disagrees with that of $\varDelta - \varDelta_{c}$.
This fact, which is one of the characteristics of the infinite-randomness fixed point, leads to a difference between the two exponents.

\subsection{Spin glass in two dimensions}

We next investigate the Ising spin glass in transverse fields, where the sign of couplings $J_{ij}$ can take both positive and negative values.
In our calculations, $J_{ij}$ is independently governed by the Gaussian distribution $P(J_{ij}) = \exp [-(J_{ij} - J_{0})^{2}/2]/\sqrt{2 \pi}$, where the variance is set equal to $1$.
We control the average $J_{0}$ of the distribution and the uniform transverse field $\Gamma$.

Let us draw the phase boundaries.
Which phase the system lies on is determined as follows.
The paramagnetic phase and the ordered phases, namely the ferromagnetic phase and the spin glass phase, are distinguished as in the case of the random ferromagnet.
Specifically, if the average of $\ln \Gamma - \ln J$ after $15$ renormalization-group transformations is larger than its initial value, the system is regarded as being in the paramagnetic phase.
Otherwise, the ferromagnetic or spin glass phase is realized.
The boundary between the ferromagnetic phase and the spin glass phase is drawn by the following rule concerning the value of $[J]^{2}/V_{J}$, where $[J]$ denotes the average of $J_{ij}$ and $V_{J}$ expresses the variance of $J_{ij}$.
We determine that the system is in the ferromagnetic phase if $[J]^{2}/V_{J}$ is larger than $1$ after $15$ renormalization-group transformations.
Otherwise, it is in the spin-glass phase.
We take $30$ samples in this calculation.

\begin{figure}
\begin{center}
\includegraphics[width=7cm,clip]{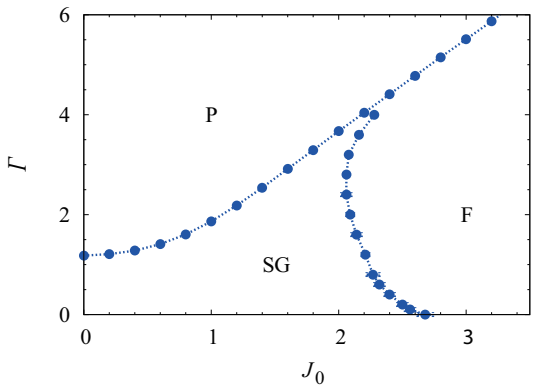}
\caption{Phase diagram of the two-dimensional spin glass in transverse fields.
The horizontal axis $J_{0}$ expresses the average of the Gaussian distribution of couplings $J_{ij}$, and the vertical axis is for the magnitude of the transverse field $\Gamma$.
The symbols, P, SG, and F, denote the paramagnetic, spin glass, and ferromagnetic phases, respectively.}
\label{fig:2DSGboundary}
\end{center}
\end{figure}

The resulting phase boundaries are depicted in Fig.~\ref{fig:2DSGboundary}.
The result is not quantitatively in precise agreement with a previous study~\cite{Rieger} on the transition point along the line $J_{0} = 0$, or more specifically, $\Gamma_{c} = 1.183(3)$ in our result and $\Gamma_{c} = 0.608(4)$ in \cite{Rieger}.
It is nevertheless important that a definite phase diagram has been obtained, especially with a spin-glass phase, by the present simple renormalization group with block-spin transformations.
This result implies that our method properly reflects the effect of frustration, which is one of the most essential features of spin glasses.

\begin{figure}
\begin{center}
\includegraphics[width=7.5cm,clip]{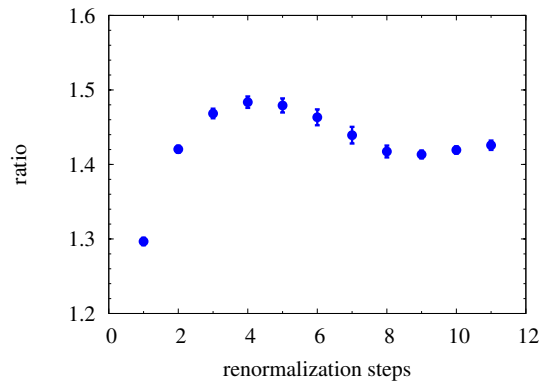}
\caption{Ratio of the square root of variance of $\ln \Gamma - \ln J$ after a renormalization to that before the renormalization in the two-dimensional spin glass in the transverse field $\Gamma = 1.195$.}
\label{fig:2DSGvariance}
\end{center}
\end{figure}

The ratio of $\sqrt{V}$ after a renormalization to that before the renormalization is observed also in this spin glass model and is plotted in Fig.~\ref{fig:2DSGvariance}, which shows the existence of an infinite-randomness fixed point in the present system.
This result supports a conjecture about the possible existence of an infinite-randomness fixed point in spin glasses inferred indirectly (i.e. without explicit calculations) from the strong disorder renormalization-group approach to the random ferromagnetic system~\cite{Motrunich}.
Moreover, our result leads to a difference between the two exponents $\nu$ and $\nu_{\mathrm{typ}}$ in the definition (\ref{eq:xityp}), as in the random ferromagnetic case.

\begin{figure}
\begin{center}
\includegraphics[width=7.5cm,clip]{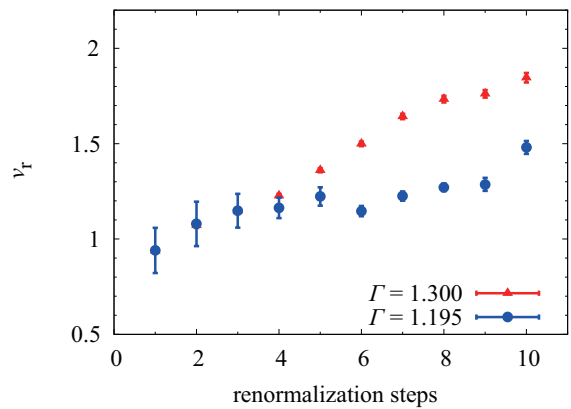}
\caption{Running exponent $\nu_{\mathrm{r}}$ calculated from Eq.~(\ref{eq:2Dnurelation}) in the two-dimensional spin glass in transverse fields.
The result in the case of $\Gamma = 1.195$, where the system is close to the critical point, has a plateau.}
\label{fig:2DSGnu}
\end{center}
\end{figure}

We next calculate the critical exponent $\nu$ for the average correlation length.
We focus on the transition point for $J_{0} = 0$.
As in the case of a random ferromagnet, the estimation is executed slightly away from the actual transition point $\Gamma = 1.183(3)$ to avoid instabilities.
Specifically, the initial transverse field is set equal to $\Gamma = 1.195$.
The running exponent $\nu_{\mathrm{r}}$ calculated in the same way as in the random ferromagnet is shown in Fig.~\ref{fig:2DSGnu}.
To compare the critical region with another case away from criticality, results with the initial transverse field $\Gamma = 1.300$ are also plotted there.
We find a difference between the two cases that the value of $\nu_{\mathrm{r}}$ for $\Gamma = 1.195$ reaches a plateau after four transformations and the value increases again after ten transformations.
Since this behavior corresponds to the plateau in the random ferromagnet, we accordingly evaluate $\nu$ with values in this region (from 4 to 9 in the horizontal axis),
\begin{equation}
\nu = 1.21(9).
\end{equation}
For comparison, if we use the values in the same region as in the case of the random ferromagnet (from 4 to 7 on the horizontal axis), we have $\nu = 1.19(8)$.

In addition to having an infinite-randomness fixed point, the value of $\nu$ estimated by our method for the spin glass is in good agreement with that for the random ferromagnet [Eq.~(\ref{eq:2Dnu})].
Although the quantitative reliability of our method in spin glasses may not be perfect as suggested in the determination of phase boundaries, the above-mentioned agreement may suggest that the random ferromagnetic Ising model in transverse fields and the Ising spin glass in transverse fields would belong to the same universality class.

\section{Three dimensions}{\label{sec:3D}} 

\subsection{Generalization to the three-dimensional models}

\begin{figure}
  \begin{center}
  \includegraphics[width=8.5cm,clip]{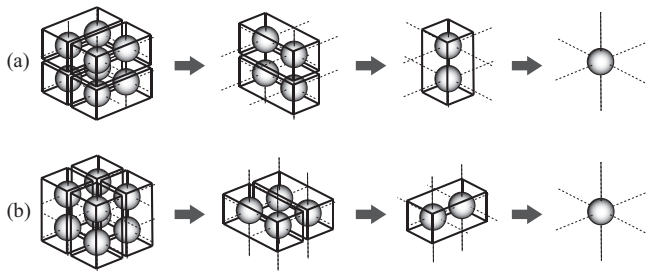}
  \caption{Three steps of renormalization in three dimensions in the (a) regular order and the (b) reverse order.}
  \label{fig:3DRG}
  \end{center}
\end{figure}

Let us next generalize our renormalization-group scheme to three dimensions.
The transformations are iterated in the horizontal, vertical, and third directions consecutively (Fig.~\ref{fig:3DRG}).
We define the coupling constants between spin $(i,j,k)$ and the neighboring spin along the third direction $(i,j,k+1)$ as $J_{t(i,j,k)}$, and similarly for $J_{h(i,j,k)}$ and $J_{v(i,j,k)}$ for the horizontal and vertical directions, respectively.

In the first step of renormalization (in the horizontal direction) the parameters change as 
\begin{equation}
\tilde{J}_{h(i,j,k)} = \frac{J_{h(2i+1,j,k)} J_{h(2i+2,j,k)}}{\sqrt{J_{h(2i+2,j,k)}^{2} + \Gamma_{(2i+2,j,k)}^{2}}},
\end{equation}
\begin{equation}
\begin{split}
\tilde{J}_{v(i,j,k)} =& \frac{J_{h(2i,j,k)}}{\sqrt{J_{h(2i,j,k)}^{2} + \Gamma_{(2i,j,k)}^{2}}} \\
& \times \frac{J_{h(2i,j+1,k)}}{\sqrt{J_{h(2i,j+1,k)}^{2} + \Gamma_{(2i,j+1,k)}^{2}}} \\
& \times J_{v(2i,j,k)} + J_{v(2i+1,j,k)}, 
\end{split}
\end{equation}
\begin{equation}
\begin{split}
\tilde{J}_{t(i,j,k)} &= \frac{J_{h(2i,j,k)}}{\sqrt{J_{h(2i,j,k)}^{2} + \Gamma_{(2i,j,k)}^{2}}} \\
& \times \frac{J_{h(2i,j,k+1)}}{\sqrt{J_{h(2i,j,k+1)}^{2} + \Gamma_{(2i,j,k+1)}^{2}}} \\
& \times J_{t(2i,j,k)} + J_{t(2i+1,j,k)}, 
\end{split}
\end{equation}
\begin{equation}
\tilde{\Gamma}_{(i,j,k)} = \frac{\Gamma_{(2i,j,k)} \Gamma_{(2i+1,j,k)}}{\sqrt{J_{h(2i,j,k)}^{2} + \Gamma_{(2i,j,k)}^{2}}}.
\end{equation}
Note that the coupling constants of the vertical and third directions are changed under the same
rule.
Carrying out analogous transformations in the vertical direction and then in the third direction after this first step, we obtain the parameters of the system renormalized in the three directions.

This scheme, however, yields anomalous results due to the imbalanced treatment of three directions.
As the transformations for three directions are iterated, the magnitude of coupling constants of the third direction $J_{t}$ rapidly becomes small in comparison with those of the other two directions despite the fact that the original system has no anisotropy.
The scheme thus needs some corrections.

To resolve the anisotropy problem, we further renormalize the system in the reverse order, along the third, vertical, and then horizontal directions after the regular order described above (Fig.~\ref{fig:3DRG}).
This set of six steps, the regular order and then the reverse order, is regarded as a single transformation of scaling factor $4$.
This modified procedure succeeds in rendering virtually isotropic results.
This is the same process as introduced in the previous study for the non-random system~\cite{Miyazaki}.

Numerical calculations are implemented in the same way as in the two-dimensional case.
We generate a pool containing $N$ couplings in the horizontal, vertical, and third directions, respectively, and $N$ fields, where $N$ is $10^{6}$.
The cluster method is used also in three dimensions with the cluster of size $L \times L \times L$ and $L$ is set equal to $5$.
We repeat the (six-step) renormalization-group transformations five times and $100$ samples have been run.

\subsection{Random ferromagnet in three dimensions}

We first examine the random ferromagnetic Ising model in transverse fields on the cubic lattice.
The initial distribution of couplings $J$ in the pool is $p(J) = \theta (J) \theta (1 - J)$, which is identical to the two-dimensional case.
We control the magnitude of the uniform field $\Gamma$ as the initial condition.

We estimate the transition point by a comparison of the initial value of $\varDelta$ and the final value of $\varDelta$ after five transformations.
The result is $\Gamma = 1.266(2)$.

\begin{figure}
\begin{center}
\includegraphics[width=7.5cm,clip]{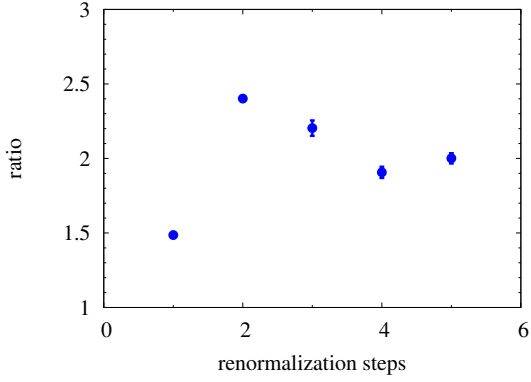}
\caption{Ratio of the square root of the variance  of $\ln \Gamma - \ln J$ after a renormalization to that before the renormalization in the random ferromagnetic Ising model in transverse fields in three dimensions.
The initial transverse field is $1.269$.}
\label{fig:3Dvariance}
\end{center}
\end{figure}

The ratio of the square root of the variance $V$ of $\ln \Gamma - \ln J$ after a renormalization to that before the renormalization is plotted in Fig.~\ref{fig:3Dvariance}.
The stability of the ratio through renormalization is not obvious, but at least the result demonstrates the existence of an infinite-randomness fixed point.

\begin{figure}
\begin{center}
\includegraphics[width=7.5cm,clip]{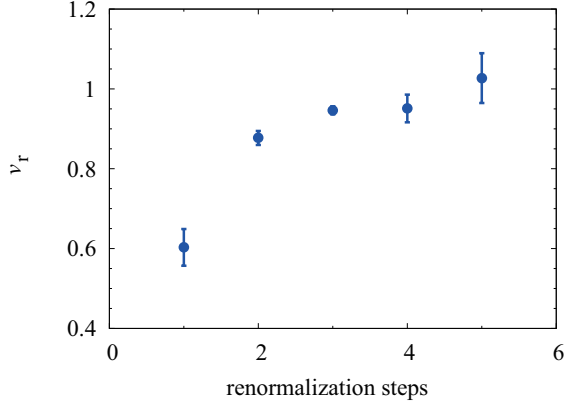}
\caption{Running exponent $\nu_{\mathrm{r}}$ calculated by Eq.~(\ref{eq:3Dnurelation}) in the random ferromagnetic Ising model in transverse fields in three dimensions.}
\label{fig:3Dnu}
\end{center}
\end{figure}

The critical exponent $\nu$ for the average correlation length is also estimated from the values of the running exponent $\nu_{\mathrm{r}}$ computed from the relation
\begin{equation}
\frac{\varDelta^{i+1} - \varDelta_{c}^{i+1}}{\sqrt{V^{i+1}}} = 4^{1/\nu_{\mathrm{r}}} \frac{\varDelta^{i} - \varDelta_{c}^{i}}{\sqrt{V^{i}}}, \label{eq:3Dnurelation}
\end{equation}
where we regard the average of $\varDelta^{i}$ for the initial transverse fields $\Gamma = 1.266$ and $1.267$ as $\varDelta_{c}^{i}$.
We can find a plateau in the result of $\Gamma = 1.269$ most clearly near the critical point (Fig.~\ref{fig:3Dnu}).
Evaluating $\nu$ from the values of $\nu_{\mathrm{r}}$ on the plateau (from 2 to 4 on the horizontal axis), we have
\begin{equation}
\nu = 0.92(4), \label{eq:3Dnu}
\end{equation}
which is rather close to $0.97(5)$ of the corresponding result by the strong-disorder renormalization group~\cite{Kovacs2, Kovacs3}.

\subsection{Spin glass in three dimensions}

Let us apply our renormalization-group scheme to the spin glass in transverse fields on the cubic lattice.
As in the case of the square lattice, we use the Gaussian distribution $P(J_{ij}) = \exp [-(J_{ij} - J_{0})^{2}/2]/\sqrt{2 \pi}$ and control the average $J_{0}$ of the distribution and the uniform transverse field $\Gamma$.

\begin{figure}
\begin{center}
\includegraphics[width=7cm,clip]{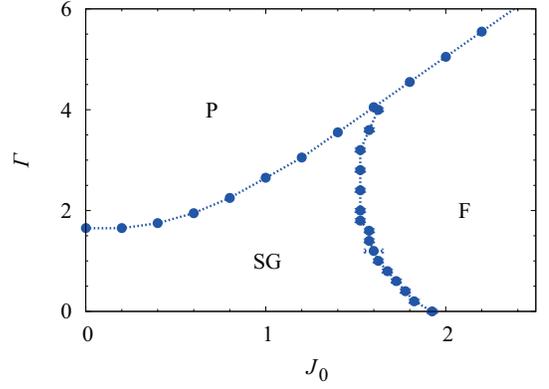}
\caption{Phase diagram of the three-dimensional spin glass in transverse fields.
The horizontal axis $J_{0}$ expresses the average of the Gaussian distribution of couplings $J_{ij}$, and the vertical axis expresses the magnitude of the transverse field $\Gamma$.
The symbols, P, SG, and F, denote the paramagnetic, spin glass, and ferromagnetic phases, respectively.}
\label{fig:3DSGboundary}
\end{center}
\end{figure}

We first draw phase boundaries under the same rule as in two dimensions, but phases are determined after five transformations in three dimensions.
The resulting boundaries are depicted in Fig.~\ref{fig:3DSGboundary}.
It is remarkable that our simple renormalization-group method verifies the existence of the spin-glass phase.

\begin{figure}
\begin{center}
\includegraphics[width=7.5cm,clip]{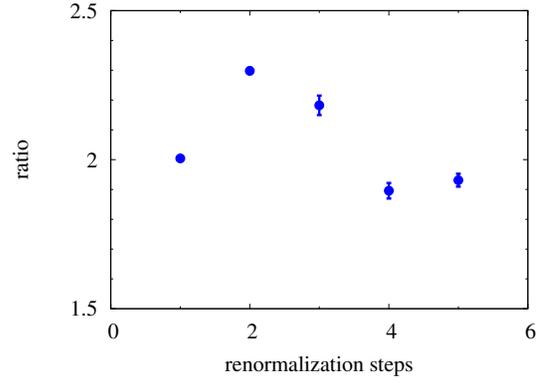}
\caption{Ratio of the square root of the variance of $\ln \Gamma - \ln J$ after a renormalization to that before the renormalization in the three-dimensional spin glass in transverse fields with $\Gamma = 1.619$.}
\label{fig:3DSGvariance}
\end{center}
\end{figure}

\begin{figure}
\begin{center}
\includegraphics[width=7.5cm,clip]{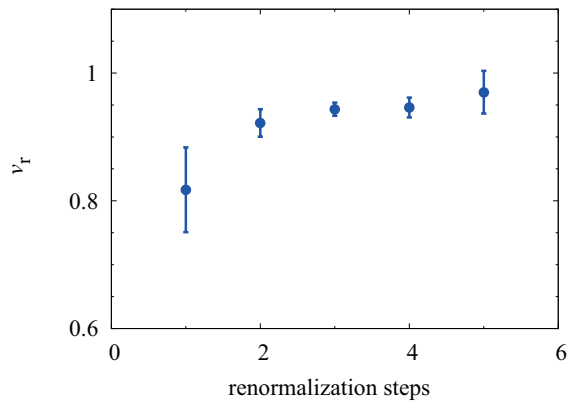}
\caption{Running exponent $\nu_{\mathrm{r}}$ calculated from Eq.~(\ref{eq:3Dnurelation}) in the three-dimensional spin glass in transverse fields with $\Gamma = 1.619$.}
\label{fig:3DSGnu}
\end{center}
\end{figure}

Next, critical properties are investigated.
Specifically, we treat the critical point for $J_{0} = 0$.
The existence of an infinite-randomness fixed point is confirmed by the observation of the change of $\sqrt{V}$ through the renormalization-group transformations (Fig.~\ref{fig:3DSGvariance}).
The running exponent $\nu_{\mathrm{r}}$ is also calculated to obtain the critical exponent $\nu$.
We use the average of $\varDelta^{i}$ for the initial transverse fields $\Gamma = 1.613$ and $1.612$ as $\varDelta_{c}^{i}$.
A plateau is clearly recognized in the result of $\Gamma = 1.619$ (Fig.~\ref{fig:3DSGnu})
and we determine $\nu$ by the values of $\nu_{\mathrm{r}}$ on the plateau (from 2 to 4 on the horizontal axis),
\begin{equation}
\nu = 0.94(3).
\end{equation}
This value is close to that of the random ferromagnetic model in three dimensions [Eq.~(\ref{eq:3Dnu})].

\section{conclusion}{\label{sec:conclusion}}

We have studied the random transverse-field Ising models in finite dimensions by the real-space renormalization-group method introduced in previous studies~\cite{Fernandez, Miyazaki}.
Our method reproduces exact results for the transition point and critical exponent $\nu$ in one dimension in spite of the existence of randomness.
Moreover, our generalization of the method to higher dimensions has been shown to be effective not only in the pure model as shown in~\cite{Miyazaki} but also in the random model.
In fact, we have obtained the values of the critical exponent $\nu$ in the two- and three-dimensional random ferromagnetic Ising models in transverse fields close to those from the strong-disorder renormalization-group approach~\cite{Kovacs, Kovacs2, Kovacs3}.

Most remarkable results in our study concern two- and three-dimensional spin glasses in transverse fields.
It should be emphasized that in this study the phase diagrams have been drawn (qualitatively) for the finite-dimensional spin glasses in transverse fields by analytical methods.
In particular, the existence of a spin-glass phase has been confirmed.
Also, we have established the existence of infinite-randomness fixed points from indefinite amplifications of the distribution of parameters on a logarithmic scale.
This observation supports the conjecture inferred from the case of a random ferromagnet under the strong-disorder renormalization group~\cite{Motrunich}, but is in conflict with a relatively old numerical study~\cite{Rieger}.
Furthermore, the estimated exponent $\nu$ in spin glasses in transverse fields is very close to that of the corresponding random ferromagnet.
Thus, one may naturally expect that these models belong to the same universality class, which is also a highly non-trivial result.
Nevertheless, further investigations are needed to settle this issue because we have to establish the quantitative reliability of the method developed here.

The validity of our method is not readily obvious, because it drops higher-energy eigenstates in the block Hamiltonians.
One may nevertheless be allowed to expect that the consistencies of our results with previous studies, wherever applicable, would justify our procedures as a method to extract the essential features of random quantum systems.
It is necessary to establish methods to evaluate other critical exponents to reinforce the quantitative reliability of the present method.

\begin{acknowledgments}
RM is grateful for the financial support from the Global Center of Excellence Program by MEXT, Japan through the ``Nanoscience and Quantum Physics" Project of the Tokyo Institute of Technology and acknowledges the financial support provided through the Research Fellowship of the Japan Society for the Promotion of Science.
\end{acknowledgments}


\end{document}